\documentclass[conference]{IEEEtran}
\IEEEoverridecommandlockouts

\usepackage{cite}
\usepackage[cmex10]{amsmath}
\usepackage{amssymb}
\usepackage{graphicx}
\usepackage{epstopdf}
\usepackage{graphics}
\usepackage{pdfpages}
\usepackage{float}
\usepackage{changes}
\usepackage{multirow}
\usepackage{mathtools}

\usepackage{tabularx}
\usepackage[utf8]{inputenc}
\usepackage{array}
\usepackage{wrapfig}
\ifCLASSINFOpdf
\else
\fi
\usepackage[cmex10]{amsmath}
\usepackage{epsfig} \ifCLASSOPTIONcompsoc
\usepackage[caption=false,font=normalsize,labelfont=sf,textfont=sf]{subfig}
\else
\usepackage[caption=false,font=footnotesize]{subfig}
\fi
\usepackage{stfloats}
\def\BibTeX{{\rm B\kern-.05em{\sc i\kern-.025em b}\kern-.08em
    T\kern-.1667em\lower.7ex\hbox{E}\kern-.125emX}}
\begin{document}
\title{Subcarrier Number  and Indices-Based Key Generation for Future Wireless Networks \thanks{This work has been submitted to the IEEE for possible publication. Copyright may be transferred without notice, after which this version may
no longer be accessible.}}

\author{
\IEEEauthorblockN{Haji M. Furqan\IEEEauthorrefmark{1}, Jehad M. Hamamreh \IEEEauthorrefmark{2} and H\"{u}seyin Arslan\IEEEauthorrefmark{1}\IEEEauthorrefmark{3}}\\
\IEEEauthorblockA{\IEEEauthorrefmark{1}Department of Electrical and Electronics Engineering, Istanbul Medipol University, Istanbul, 34810 Turkey\\}

\IEEEauthorblockA{\IEEEauthorrefmark{2}Department of Electrical Engineering, Antalya Bilim University\\}
\IEEEauthorblockA{\IEEEauthorrefmark{3}Department of Electrical Engineering, University of South Florida, Tampa, FL, 33620 }
Email: \{haji.madni@medipol.edu.tr, jehad.hamamreh@antalya.edu.tr, huseyinarslan@medipol.edu.tr\}}
\maketitle
\begin{abstract} 
\textcolor{black}{Physical layer key generation from the wireless channel is an emerging area of interest to provide confidentiality and authentication. One of the main challenges in this domain is to increase the length of the secret key while maintaining its randomness and uniformity. In this work, new dimensions for wireless channel-based key generation are proposed for orthogonal frequency division multiplexing (OFDM) systems. The novel perspective of the proposed work lies in the generation of key bits not only from the magnitudes of OFDM subchannels as it has conventionally been done but also from the number and positions of those subchannels whose channel gains are above the mean of the respective subblock. The effectiveness of the proposed algorithms is evaluated in terms of key generation rate and key mismatch rate. Additionally, a statistical test suite offered by the National Institute of Standards and Technology is used to evaluate the randomness of the generated key bits. It is shown in the simulation results that the involvement of the proposed dimensions can double the key generation rate compared to conventional algorithms.}
\end{abstract}
\section{Introduction}
\textcolor{black}{Due to the broadcast nature of wireless communication, it is vulnerable to eavesdropping and spoofing attacks \cite{9336039}. The conventional encryption-based solutions to tackle such attacks may not be feasible in 5G and beyond heterogeneous wireless networks due to the complexity of generation, management, and sharing of secret keys among different entities. In order to resolve such issues, physical layer key generation provides an alternative to encryption-based methods by exploiting the channel reciprocity property as a common source of randomness between communicating nodes. Also, the rich multipath environment will ensure that the attacker will experience an independent channel if it is half-wavelength apart from legitimate nodes. Besides, there is no need for any infrastructure to distribute physical layer keys \cite{8883129}.}

\textcolor{black}{Physical layer key generation in multi-carrier systems such as orthogonal frequency division multiplexing (OFDM) has attracted significant interest due to increasing security and pricey concerns. The major works in this direction are based on the exploitation of received signal strength (RSS), amplitude and phase of the channel impulse response (CIR)/channel frequency response (CFR), and other feedback mechanisms \cite{7393435}. The authors in \cite{Liu2013k} exploit RSS for channel-based key generation in the OFDM system for both indoor and outdoor environments.  In \cite{Zhang2016}, subcarriers’ channel responses are utilized in OFDM for secret key generation. Similarly, an experimental setup for key generation from subcarriers’ channel responses in OFDM is presented in \cite{Cheng2017k}. In \cite{Liu2012}, multiple independent phases are quantized for secret key generation in a multi-carrier system.  Likewise, the authors make use of phase change of the time-varying channel frequency response corresponding to OFDM subcarriers in \cite{Wang2011}. In \cite{Wu2013a}, OFDM subcarriers are exploited along with precoding matrix indices for secret key generation. }

\textcolor{black}{The above-mentioned conventional key generation-based algorithms are effective methods for key generation. However, they have certain limitations in terms of generating longer keys while keeping the randomness and uniformity of the key bits \cite{7393435}. In order to enhance the key generation rate, we proposed novel dimensions of key generation for multicarrier communication systems in \cite{9201305}, where the indices of sub-channel corresponding to best channel gains are exploited for key generation along with the amplitude. We further build on top of
\cite{9201305} by proposing the use of number, position, and amplitude of subchannels in OFDM, which further enhances the overall key rate compared to \cite{9201305}.} 

\textcolor{black}{
The main contributions of the proposed work are as follows.
\begin{itemize}
\item New dimensions of key generation are proposed, where the number of subchannels in each subblock of the OFDM block along with their position and amplitudes, whose channel gains are above the mean of the subblock, are exploited for a key generation. The proposed methods are inspired by the concept of OFDM with subcarrier number modulation (SNM) \cite{8362748}. Particularly, the whole OFDM block is divided into subblocks, where the number of the subcarrier in each subblock is used similar to OFDM SNM. However, here the number of subcarriers is not chosen based on the data to convey information as in OFDM-SNM but rather the number and positions, of those subcarriers in each subblock whose gains are above the mean of channel gains of each subblock, are used for key generation. 
\item The proposed approaches can be used to provide confidentiality and authentication against eavesdropping and spoofing attacks in any multicarrier system such as OFDM, OFDM-SNM \cite{8417419}, and OFDM-index modulation. Besides, there is a potential to extend this work to other domains such as space, time, and code.  
\item Several performance metrics, including key generation rate (KGR) and key mismatch rate (KMR) are also evaluated to get valuable insights about the proposed algorithm. Moreover, a statistical test suite offered by the National Institute of Standards and Technology (NIST) is adopted to check the randomness of generated key bits.
\end{itemize}}
\section{System Model}
\begin{figure}[h]
	\centering
	\includegraphics[scale=0.6]{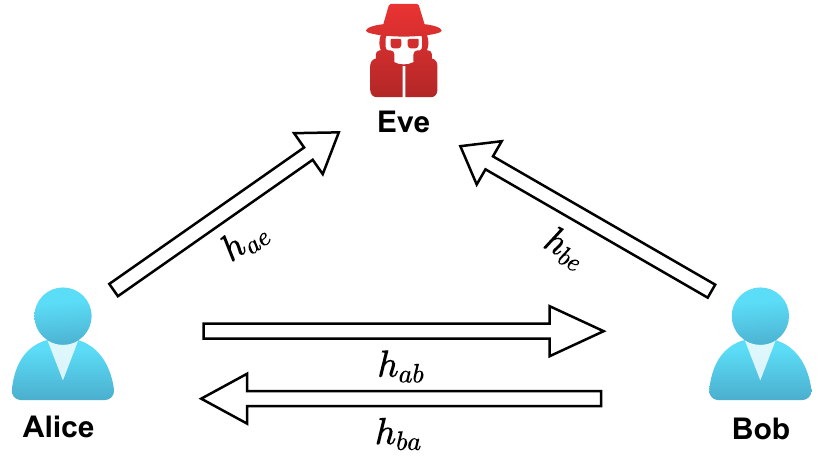}
	\caption{A simplified system model for the considered security algorithm.} 
	\label{fig:systemmod}
\end{figure}

\textcolor{black}{A single-input single-output (SISO)-OFDM system is assumed with a 
time division duplexing (TDD) configuration that consists of single antenna legitimate nodes (Alice and Bob) that want to communicate in the presence of a passive single antenna eavesdropper (Eve) as presented in Fig. \ref{fig:systemmod}. We assume slow varying Rayleigh fading channel with $L$ exponentially decaying taps between Alice-Bob ($\mathbf{h_{ab}} \in \mathbb{C}^{[1 \times L]}$) and Alice-Eve $(\mathbf{h_{ae}} \in \mathbb{C}^{[1 \times L]})$. Channel reciprocity property is assumed between the node pair such that Alice can estimate the channel from it to Bob ($\mathbf{h_{ab}}$) based on channel from Bob to it ($\mathbf{h_{ba}}$). Moreover, it is also assumed that the legitimate transmitter has no knowledge about the channel state information of the eavesdropper.}

\textcolor{black}{At the transmitter, the frequency domain domain OFDM symbols (pilots) $\mathbf{{s}}=\begin{bmatrix} s_{0} & s_{1} &...& s_{N-1}\end{bmatrix}^{\textrm T} \in \mathbb{C}^{[N \times 1]}$ are passed through IFFT process $\mathbf{F}^{\textrm H} \in \mathbb{C}^{[N \times N]}$ to map them to orthogonal subcarriers, where $N$ represents the number of frequency domain complex data symbols (pilots) and $\mathbf{F}$ is the discrete Fourier transform matrix. Afterward, cyclic prefix (CP) appending matrix $\mathbf{C} \in \mathbb{R}^{[(N+L) \times N]}$ is employed to insert a CP of length $L$. Then, the resultant signal in time domain is transmitted through channel and reaches Bob and Eve.
 The received signal at Bob/Eve in frequency domain after removing CP and applying FFT, can be given as
\begin{eqnarray}  
\mathbf{y_{b/e}}&=&\mathbf{F}\mathbf{D}\left( \mathbf{H_{ab/ae}} \mathbf{C}\mathbf{F}^{\textrm H}\mathbf{s}+\mathbf{z_{ab/ae}}\right), \\
&=& \mathbf{H_{ab/ae}^{f}}\mathbf{s}+\mathbf{\hat{z}_{ab/ae}}, \label{bobRX}
\end{eqnarray} 
where $\mathbf{D} \in \mathbb{R}^{[N \times (N+L)]}$ is CP removing matrix, $\mathbf{z_{b/e}}$ and $\mathbf{\hat{z}_{ab/ae}}$ represent the zero-mean complex additive white Gaussian noise (AWGN) and its Fourier transform, respectively, with variance of $\sigma_{ab/ae}^2$ at Bob/Eve. On the other hand, $\mathbf{H_{ab/ae}} \in \mathbb{C}^{[(N+L) \times (N+L)]}$ and $\mathbf{H_{ab/ae}^{f}}=\mathbf{F}\mathbf{D} \mathbf{H_{ab/ae}}\mathbf{C}\mathbf{F}^{\textrm H} \in \mathbb{C}^{[N \times N]}$ represent the Toeplitz matrix corresponding to the channel impulse response and the diagonal matrix corresponding to the channel frequency response, respectively, with respect to Bob/Eve.}
	\begin{figure*}[h]
		\centering
		 \includegraphics[width=0.65\textwidth]{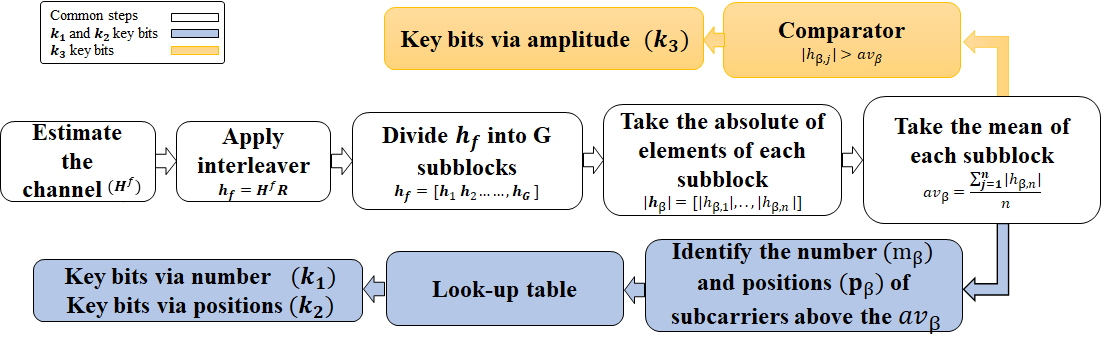}
		\caption{Proposed NIKG and NIAKG algorithms, where NIKG includes all blocks except the orange ones while NIAKG includes all blocks.}
		\label{fig:algoi}
	\end{figure*}
\section{Proposed Algorithm}\label{pro}
\textcolor{black}{This section presents the details about the proposed number and indices-based key generation (NIKG) and number, indices, and amplitude-based key generation (NIAKG) algorithms. Fig. \ref{fig:algoi} shows the summary of basic steps of the NIKG and NIAKG algorithm. }

\textcolor{black}{The basic steps for the NIKG algorithm are given below.} 
\textcolor{black}{
\begin{enumerate}
\item In the first step, to estimate the channel ($\mathbf{h}$) between Alice and Bob, they transmit pilot signals, $\mathbf{{s}}=\begin{bmatrix} s_{0} & s_{1} &...& s_{N-1}\end{bmatrix}^{\textrm T} \in \mathbb{C}^{[N \times 1]}$, to each other within the coherence time of the channel. The frequency domain estimation of channel ($\mathbf{h}$) at Alice/Bob can be represented as $\mathbf{H^{f}} \in \mathbb{C}^{[N \times N]}$, where $\mathbf{H^{f}}$ is a diagonal matrix.
\item The diagonal channel matrix at each node is multiplied with a random interleaver, $\mathbf{R}$, in order to minimize correlation between different subchannels by distributing the deep fades of subchannel uniformly over the whole OFDM block. The resultant signal can be given as $\mathbf{h_{f}}=\mathrm{diag}(\mathbf{H^{f}}\mathbf{R})$, where $\mathbf{h_{f}} \in \mathbb{C}^{[N \times 1]}$. 
\item $\mathbf{h_{f}}$ vector is sliced into $G$ subblocks with $n$ elements in each subblock, where $G=N/n$, $\mathbf{h_{f}}=[\mathbf{h}_{1},\dots,\mathbf{h}_{G}]$, $\mathbf{h_{\beta}}=[h_{\beta,1},\dots,h_{\beta,n}]$, and $\beta \in \{{1,\dots,G}\}$. 
\item The average of absolute values of elements of each subblock is calculated as $av_\beta=\frac {\sum_{j=1}^{n} |h_{\beta,j}|}{n}$.
\item Each element of $\mathbf{h_{\beta}}$ subblock is compared with mean, $av_\beta$, and the number ($m_{\beta}$) of subcarriers that are above the mean are counted for each subblock.
\item Finally, secret key bits are generated at the nodes corresponding to the number and positions of subcarriers for each subblock using Table \ref{keybits}. 
The first column of Table \ref{keybits} presents the index for  any subblock ($\beta$), the second column presents a vector $\mathbf{p_{\beta}}$ containing those subcarriers in $\mathbf{h_{\beta}}$ subblock whose channel gains are above the average value $av_\beta$, the third column depicts the number/count $m_{\beta}$ of those subcarriers, the fourth column shows the  key bits $\mathbf{k}_1$ generated by the position of subcarrier, and the fifth column presents key bits $\mathbf{k}_2$ corresponding to number/count. Here, the total number generated key bits in any OFDM block is a function of $n$, $N$, and $G$ that can be given as follows: 
${G} ( \lfloor \log_{2}\binom{n}{1} \rfloor + \log_{2} n)$.
\end{enumerate}
\begin{table} [h] 
\centering\renewcommand{\arraystretch}{1.4}
\begin{center} 
\caption{Look-up table for NIKG algorithm for $n$=4}\label{keybits}
\begin{tabular}{|c|c|c|c|c|}
\hline $\beta$ &  $\mathbf{p_{\beta}}$ & $m_{\beta}$ & $\mathbf{k}_1$ & $\mathbf{k}_2$  \\
\hline  1 & $[s_0 \thinspace0 \thinspace0 \thinspace0]^{T}$ & 1 & [0 0] & [0 1]\\
\hline  2 & $[0 \thinspace s_1 \thinspace0 \thinspace0]^{T}$ & 1 & [0 0] & [1 0]\\
\hline  3 & $[0 \thinspace0 \thinspace s_2 \thinspace0]^{T}$ & 1 & [0 0] & [1 1]\\
\hline  4 & $[0 \thinspace0 \thinspace0 \thinspace s_3]^{T}$ & 1 & [1 1] & [0 0]\\
\hline  5 & $[s_0 \thinspace s_1 \thinspace0 \thinspace0]^{T}$ & 2 & [0 1] & [0 0]\\
\hline  6 & $[s_0 \thinspace0 \thinspace s_2 \thinspace0]^{T}$ & 2 & [0 1] & [0 1]\\
\hline  7 & $[s_0 \thinspace0 \thinspace0 \thinspace s_3]^{T}$ & 2 & [0 1] & [1 0]\\
\hline  8 & $[0 \thinspace s_1 \thinspace0 \thinspace s_3]^{T}$ & 2 & [0 1] & [1 1]\\
\hline  9 & $[s_0 \thinspace s_1 \thinspace s_3 \thinspace0]^{T}$ & 3 & [1 0] & [0 0]\\
\hline 10 & $[s_0 \thinspace0 \thinspace s_2 \thinspace s_3]^{T}$ & 3 & [1 0] & [0 1]\\
\hline 11 & $[s_0 \thinspace s_1 \thinspace0 \thinspace s_3]^{T}$ & 3 & [1 0] & [1 0]\\
\hline 12 & $[0 \thinspace s_1 \thinspace s_2 \thinspace s_3]^{T}$ & 3 & [1 0] & [1 1]\\
\hline 13 & $[0 \thinspace0 \thinspace s_2 \thinspace s_3]^{T}$ & 2 & [1 1] & [0 1]\\
\hline 14 & $[0 \thinspace s_1 \thinspace s_2 \thinspace0]^{T}$ & 2 & [1 1] & [1 0]\\
\hline
\end{tabular} 
\end{center}
\end{table} 
To explain the algorithm more clearly, let's assume that we have $N=128$, $n=4$. The vector $\mathbf{h_{f}}$ corresponding to OFDM block is divided into $G=N/n=128/4=32$ subblocks. Afterward, each element of $\beta$ subblock is compared with mean, $av_\beta$, and the number of subcarriers that are above the mean is counted and their indices ($\mathbf{p_{\beta}}$) and count ($m_{\beta}$) are used for key generation using Table \ref{keybits}, where $128$ secret bits can be generated using the NIKG algorithm.}
	
\textcolor{black}{In order to further increase the key rate of the NIKG approach, a joint approach known as NIAKG is suggested. In the NIAKG approach, secret bits are generated based on the number, indices, and amplitudes of subcarriers jointly as follows. 
\begin{enumerate}
\item The initial five steps of NIAKG are similar to that of NIKG. Particularly, similar to NIKG, the estimated channel is divided into subblocks and the average value ($av_\beta=\frac {\sum_{j=1}^{n} |h_{\beta,j}|}{n}$) of elements for each subblock is calculated and each element is compared with the mean. Afterwards, key bits ($\mathbf{k}_1$ and $\mathbf{k}_2$) corresponding to each subblock are generated using Table \ref{keybits} based on $\mathbf{p_{\beta}}$ and $m_{\beta}$, respectively. 
\item In the next step, the absolute value of each subblock's element ($|h_{\beta,j}|$) is compared with the corresponding mean, $av_\beta$, in order to generate secret key bits ($\mathbf{k}_3$). More specifically, if the value of the subblock's element is greater than the average value of the subblock ($|h_{\beta,j}|>av_\beta$) then $1$ is generated; otherwise, $0$ is generated. \end{enumerate}
The total number of bits generated by NIAKG can be given as 
${G} ( \lfloor \log_{2}\binom{n}{1} \rfloor + \log_{2} n)+N$.
For example, for $N=128$ and $n=4$ the total number of secret key bits generated by the NIAKG algorithm is $256$.}

\textcolor{black}{It should be noted that the secret bits generated at legitimate and illegitimate nodes are different due to channel decorrelation assumption.}

\textcolor{black}{In order to show the effect of different parameters on key generation Table \ref{trendofnn} is presented. Particularly, Table \ref{trendofnn} shows the effect of block size (n) and number of block (G) on the number of maximum key bits generated by NIKG and NIAKG algorithms based on ${G} ( \lfloor \log_{2}\binom{n}{1} \rfloor + \log_{2} n)$ and ${G} ( \lfloor \log_{2}\binom{n}{1} \rfloor + \log_{2} n)+N$, respectively. It is observed that the maximum number of bits can be generated when $n=4$, $G=32$. Hence, $n=4$ will be used in the simulation setup.}

\begin{table} 
\centering\renewcommand{\arraystretch}{1.4}
\caption{Effect of subblock size on key bit length}\label{trendofnn}
\begin{tabular}{|l|l|l|} 
\hline
N=128 & NIKG & NIAKG  \\ 
\hline
n=4, G=32  & 128  & 256    \\ 
\hline
n=8, G=16   & 96   & 224    \\ 
\hline
n=16, G=8  & 64   & 192    \\ 
\hline
n=32, G=4  & 40   & 168    \\
\hline
\end{tabular}
\end{table}

\section{Simulation results}
\textcolor{black}{This section presents the simulation results to analyze the effectiveness of the proposed algorithms. A SISO-OFDM system consisting of $N=128$ subcarriers along with CP size of length $L$ is assumed. Based on the proposed algorithms, the OFDM subblock is divided into $G=N/n=32$ subblocks, where $n=4$. The length of the multipath Rayleigh fading channel having exponentially decaying power delay profile between any pair of the node (Alice-Bob, Alice-Eve, Bob-Eve) is assumed with $L=9$ taps \cite{cho2010mimo}. Additionally, we consider that Eve perfectly knows the proposed key generation algorithm and will employ it to generate key bit based on its channel.}
\begin{figure}[t]
	\centering
	\includegraphics[scale=0.45]{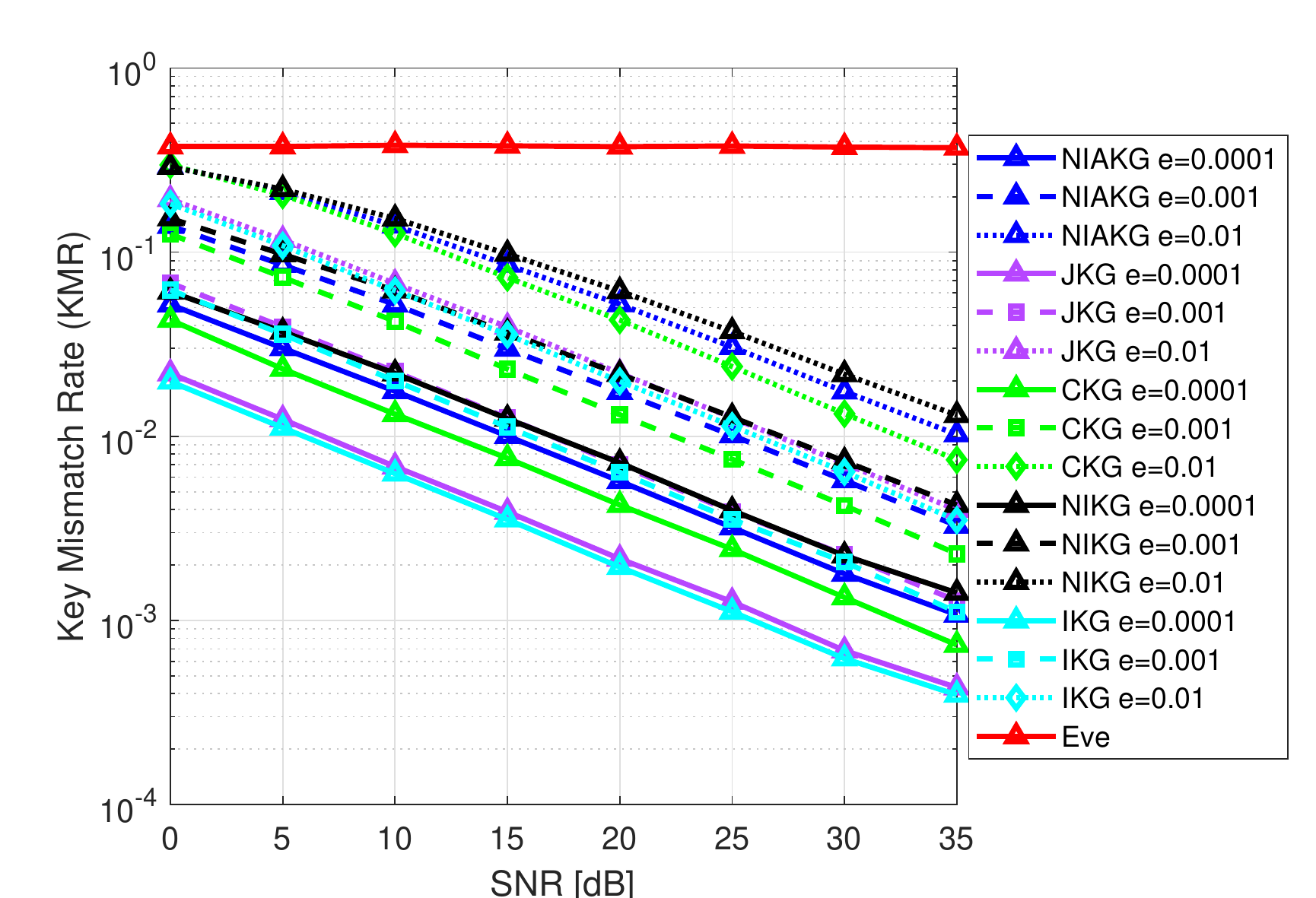}
	\caption{KMR versus SNR performances at Bob and Eve under imperfect channel reciprocity and imperfect channel estimation for NIKG, NIAKG, IKG, JKG, and CKG approaches.}
	\label{fig:KMR}
\end{figure}
\textcolor{black}{The comparison between the performances of proposed NIKG and NIAKG algorithms, indices based key generation (IKG) and joint indices based key generation (JKG) \cite{9201305}, and conventional key generation (CKG) approaches \cite{Liu2013k} are presented here. NIKG and NIAKG algorithms are already explained in section \ref{pro} while the CKG approach is based on the generation of key bits by comparing estimated channel coefficients' amplitudes with their mean \cite{Liu2013k}. IKG algorithm is based on utilization of only indices of subchannels while JKG approach exploits both indices and amplitude for key generation as explained in \cite{9201305}. KMR and KGR are used to show the effectiveness of the proposed algorithms \cite{7600974}. Additionally, a statistical test suite provided by NIST is used to evaluate the randomness of the generated key bits. Moreover, in order to show the robustness of the proposed algorithms against imperfect channel estimation and imperfect channel reciprocity, the effect of imperfections is also taken into consideration \cite{7814269}. The imperfect channel at Alice/Bob can be given as 
$\mathbf{\hat H^{f}}= \mathbf{H^f}+\mathbf{\Delta H^{f}}$. $\mathbf{H^f}$ shows the perfect channel, $\mathbf{\sigma^2}=e\times 10^{\frac{-SNR_{dB}}{10}}$ represents error variance, $e\in\mathbb{R}$ is a scale that corresponds to estimator quality, and $\mathbf{\Delta H^{f}}$ represents independent complex Gaussian noise vectors with zero mean and error variance $\mathbf{\sigma^2}$. Note that, in current work, three estimators having different qualities are considered. These qualities are incorporated by considering different values of $e$, $e \in \{0.01, 0.001, 0.0001\}$.}

\begin{figure}[h]
	\centering
	\includegraphics[scale=0.45]{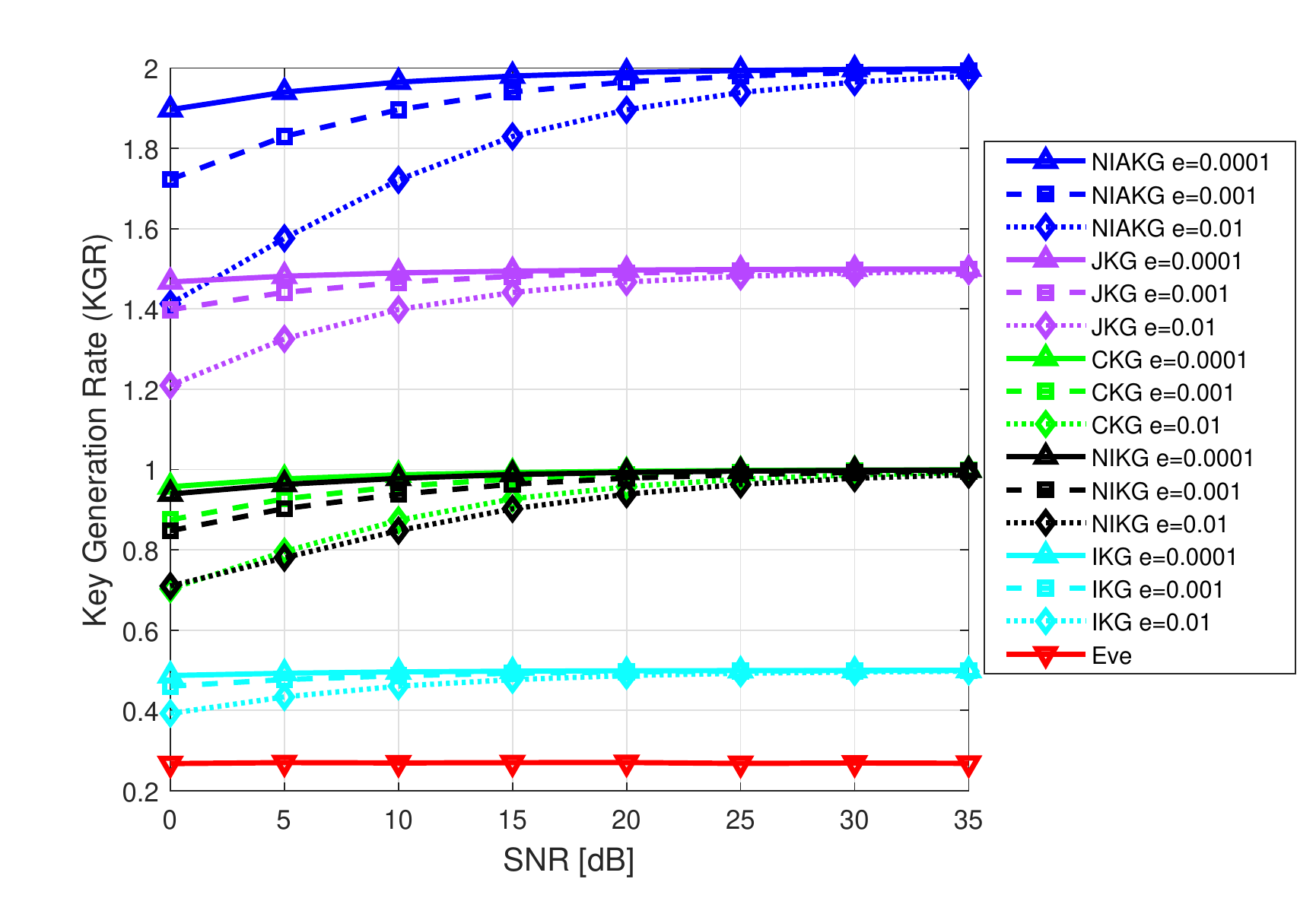}
	\caption{KGR versus SNR performances at Bob and Eve under imperfect channel reciprocity and imperfect channel estimation for NIKG, NIAKG, IKG, JKG, and CKG approaches.}
	\label{fig:KR}
\end{figure}

\textcolor{black}{Fig. \ref{fig:KMR} shows the KMR plots against signal-to-noise ratio (SNR) to show the mismatch between the generated key bits at Alice and Bob under imperfect channel assumption for NIKG, NIAKG, IKG, JKG, and CKG approaches, where independent imperfections with equal variance are assumed at them. It is clear from the figure that KMR for all algorithms decreases as the values of $e$ decrease from $0.01$ to $0.0001$ which actually corresponds to the reduction in the value of variance ($\mathbf{\sigma^2}$). It is also observed that the KMR for NIKG and NIAKG is worst compared to others (IKG, JKG, and CKG) while it is best for the IKG approach. On the other hand, the KMR performance of NIAKG is somewhere between NIKG and CKG. Furthermore, the figure also shows that there is a clear gap between KMR at legitimate node and Eve for all cases due to independent channel observation at Eve compared to observed
channel at Alice/Bob.}

\textcolor{black}{Fig. \ref{fig:KR} shows the KGR (efficiency= bits/channel coefficient) plots for NIKG, NIAKG, IKG, JKG, and CKG approaches. It is clear from the figure that the KGR performance for all approaches improves as the values of $e$ decrease from $0.01$ to $0.0001$. It is also observed that the proposed NIAKG has the best KGR performance while the KGR performance of IKG is the worst. CKG and NIKG have similar performance while the performance of JKG is in between the KGR performance of NAIKG and IKG.}
\textcolor{black}{Moreover, the joint consideration of the KMR and KGR of NIAKG and conventional CKG from Fig. \ref{fig:KMR} and Fig. \ref{fig:KR}, respectively, shows that the proposed dimensions in the NIAKG algorithm enhance the overall key rate compared to the CKG algorithm. Particularly, there is a $100\%$ increase in the key rate for the proposed NIAKG compared to the CKG algorithm with the involvement of the new dimensions for key generation, $33.33\%$ compared to JKG, $150\%$ compared to IKG algorithm.}
This subsection presents the effect of the correlation between the channel of legitimate receiver and Eve and evaluate the performance in terms of BER as a security metric.
\begin{figure}[h]
	\centering
	\includegraphics[scale=0.45]{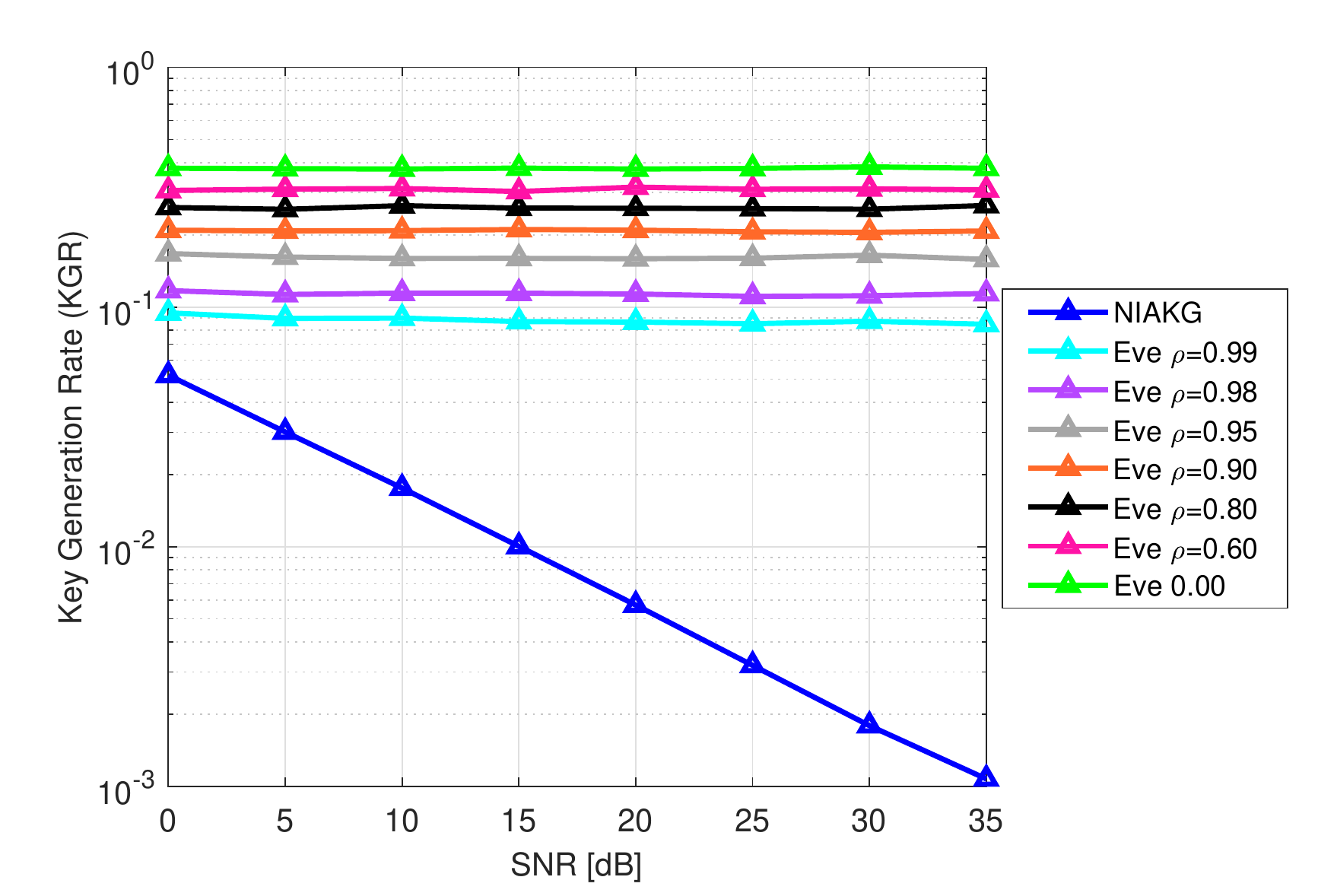}
	\caption{KGR versus SNR performance of Eve under correlation channel coefficient such as ($\rho$=0.99, 0.98, 0.95, 0.90, 0.80, 0.60, 0).
	}
	\label{fig:KRcor}
\end{figure}

Fig. \ref{fig:KRcor} show KMR versus SNR performance of Eve for the case when it has correlated channel with Alice/Bob.
The model for correlation channel assumed
in this work is $h_{e}=\rho h_{ab}+ (1-\rho) h_{i}$ \cite{8240989}, where $\rho$ is a correlation factor, $h_{e}$ is correlated channel of Eve, $h_{i}$ represents independent channel component. Particularly, KMR performance for NIAKG with $e$=0.0001 is evaluated between Alice-Bob (represented in blue color) along with Eve having correlated channel coefficient of $\rho$=(0.99, 0.98, 0.95, 0.90, 0.80, 0.60, 0). It is observed from the figure that as the value of $\rho$ increases the KMR performance improves at Eve. However, there is still a significant gap between KMR performances of legitimate node and Eve even for $\rho$=0.99.

\textcolor{black}{In order to assess the randomness of the generated key bits, a statistical test suite that is provided by NIST \cite{rukhin2001statistical} is employed. There are $15$ tests in NIST’s statistical test suite, where each of the tests assesses a specific random feature. For example, the proportion of zeros and ones are evaluated by frequency test while the DFT test is employed to detect the periodic features of the sequence. In order to check the randomness, each of test return a $P-value$ that is compared with the significance value, $\alpha \in [0.001, 0.01] $. The sequence is accepted as a random sequence if $P-value$ is greater than $\alpha$ otherwise it is assumed to be non-random. The key bitstream generated by the algorithms presented in this paper fulfills the input size requirements of $8$ of the NIST tests because the other tests need a very long sequence e.g. $10^6$. Hence, we ran $8$ tests, which still fulfill the requirements of NIST as discussed in \cite{rukhin2001statistical}. It is clear from Table \ref{nisttest} that the bitstream generated by the proposed algorithms passed the randomness test provided by the NIST suite. Particularly, the $P-values$ of our cases are above $0.01$.}

	\begin{table}[h!]\centering\renewcommand{\arraystretch}{1.7}
		\label{nist}
		\caption {NIST statistical test suite results. The $P-value$ from each test is listed below. To pass a test, the $P-value$ for that test must be
			greater than 0.01.} 
		\label{nisttest} 
		\centering
		\resizebox{.85\columnwidth}{!}{
			\begin{tabular}{|l|l|l|}
				\hline
				{Test name}  & {$P-values$ (NIKG)} & {$P-values$ (NIAKG)} \\ \hline
				{Frequency}                                                                                   & {0.75}                                                                  & {0.72}                                                                  \\ \hline
				{Block frequency}                                                                             & {0.70}                                                                  & {0.67}                                                                  \\ \hline
				{Runs}                                                                                        & {0.68}                                                                  & {0.71}                                                                  \\ \hline
				{Longest run of 1s}                                                                          & {0.60}                                                                  & {0.57}                                                                  \\ \hline
				{DFT}                                                                                         & {0.66}                                                                  & {0.63}                                                                  \\ \hline
				\multirow{2}{*}{Serial}             & \multirow{2}{*}{{\begin{tabular}[c]{@{}l@{}}0.57\\ 0.59\end{tabular}}} & \multirow{2}{*}{{\begin{tabular}[c]{@{}l@{}}0.63\\ 0.55\end{tabular}}} \\
				&                                                                                 &                                                                                                                                                        \\ \hline
				{Approx. Entropy}                                                                             & {0.50}                                                                  & {0.49}                                                                  \\ \hline
				{Cumulatic sum (forward)}                                                                  & {0.62}                                                                  & {0.59}                                                                  \\ \hline
				{Commulative sum (reverse)}                                                                   & {0.58}                                                                  & {0.56}                                                                  \\ \hline
		\end{tabular}}
\end{table}

\section{Conclusion and future directions}
\textcolor{black}{The work proposed novel dimensions for secret key generation based on the number, position, and amplitude of subchannels in each subblock of OFDM.
 Particularly, in the first step, the channel at the communicating node is converted into random order by utilizing a random interleaver. Afterward, the resultant channel response in the frequency domain is divided into small subblocks. Finally, the key bits are generated by amplitudes of individual subcarriers along with the number and positions of those subcarriers in each subblock of OFDM whose gains are above the mean of channel gains of each subblock. The proposed novel dimension of key results in the enhancement of the overall enhancement of KGR as presented in simulation results. Particularly, there is an increase of $100 \%$ in key rate as presented by NIAKG compared to CKG due to the involvement of a new dimension, $33.33\%$ compared to JKG, and $150\%$ compared to the IKG algorithm. For future work, different variations of the proposed algorithm in the other domains such as space, time, and code will be exploited.}
\section{Acknowledgment}
The work of Haji M. Furqan was supported by the HISAR Lab at TUBITAK BILGEM, Gebze, Turkey. 
Author is thankful to Muhammad Sohaib J. Solaija, Sheema, Nann, and Liza afeef for their constructive comments.



\end{document}